\documentclass[seceq]{ptptex}
\def\vc#1{\mbox{\boldmath $#1$}}
\usepackage{graphicx}

\title{New Treatment of Resonances with Bound State Approximation by Using Pseudo Potential}

\author{Yasuro \textsc{Funaki}$^1$, Hisashi \textsc{Horiuchi}$^1$ and Akihiro \textsc{Tohsaki}$^2$
}

\inst{$^1$Department of Physics, Kyoto University, Kyoto 606-8502, 
Japan \\  $^2$Suzuki Corporation, 46-22 Kamishima-cho Kadoma, Osaka 571-0071 Japan}

\abst{We propose a new approach to extract the wave functions of resonances by the bound state approximation which gives the mixed states of the resonance components and the continuum ones. In our approach, on the basis of the method of analytic continuation in the coupling constant (ACCC), we construct Pad\'e rational function by adopting the positive energies as well as the negative ones. We report the result of the application of this new method to the second $2^+$ state of $^{12}$C which was studied with the ACCC method in our previous work. It is found that the resonance parameters obtained by the ACCC method are well reproduced by the new method. Some advantages over the ACCC method are also shown.}

\begin{document}

\maketitle

\section{Introduction}
 In order to study broad resonance states, it is well known that the bound state approximation does not work well. Among the methods which can treat resonance states, we recently see many studies which are based on the ACCC method\ \cite{accc,TSVL,aoyama} and complex scaling method\ \cite{csm,KLGK}. In the case of the ACCC method, however, one can only calculate the energy and width of the resonance but to get the resonance wave function is almost impossible. On the other hand, in the case of the complex scaling method, one can in principle get the wave function in addition to the resonance energy and width, which is however not easy in practice. Moreover, the wave function which is to be obtained by the so-called back rotation applied to the complex-rotated wave function is the so-called Gamow state which is exponentially diverging in the outside spatial region and is not easy to handle.

 The purpose of this paper is to propose an extended version of bound state approximation to the description of resonances which provides us with a new method of the calculation of the energy, the width and the wave function of the resonance. In our extended version of the bound state approximation, we first perform the calculation of usual bound state approximation and obtain the energy eigen-states.  Since the energy eigen-states consist of the mixture of the resonance wave function and the continuum state components, we then extract the resonance component by superposing energy eigen-states. For this procedure we need to know approximately how is the mixing between resonance states and continuum states, and this knowledge is obtained by using the pseudo potential just like in the ACCC approach. The difference of our method from the original ACCC method is that we can also keep the coupling constant of the pseudo potential so weak that the resonance state does not become the bound state, while in the original ACCC method it is inherently necessary to use the strong coupling constant which converts the resonance state to the bound state. The first part of our method, namely the extraction of the resonance state component from energy eigen-states, was already utilized recently by the present authors in Ref.\ \cite{funakib}. Here in this paper, we give a full explanation of our method and the calculational method of the resonance width. 

 The content of this paper is as follows. In \S\ref{treat}, we explain our new treatment of the resonance state in the bound state approximation with the aid of the pseudo potential. It consists of two parts. The first part is the extraction of the resonance wave function from the energy eigen-state wave functions obtained by the bound state approximation. The second part is the calculation of the resonance width.  In \S\ref{appli}, we show the numerical example of the application of this new method. The application is done to the second $2^+$ state of $^{12}$C which was recently observed at $2.6\pm 0.3$ MeV above the $3\alpha$ breakup threshold with $\Gamma = 1.0 \pm 0.3$ MeV \cite{Itoh}. Since this state was investigated by using the ACCC method in the previous paper by the present authors \cite{funakib}, we here compare the results of the new ACCC and the previous one. We show that the complex energy function obtained by our method is in good agreement with the one of the original ACCC method in a wide coupling constant region, and therefore our method gives the energy and width which well agree with those obtained by the ACCC method. \S\ref{conc} is for conclusion.

\section{New Treatment of Resonances with Bound State Approximation by the Aid of Pseudo Potential}\label{treat}

\subsection{Extraction of resonance wave function}\label{extraction}

 The calculation of the resonance state in bound state approximation is usually made by diagonalizing the Hamiltonian by using a finite number of basis wave functions which are square integrable.  The calculated positive energy eigen-states which are linear combinations of basis wave functions are divided into resonance states and continuum states.  There are many practical methods to make this division, namely the selection of the resonance states among positive energy eigen-states. One method is to introduce an attractive pseudo potential $V$ which is added to the original Hamiltonian $H$, 
\begin{equation}
  H'(\delta) = H + \delta \times V, \label{hamil}
\end{equation}
 where $\delta$ is a coupling constant to vary the strength of the pseudo potential.  We diagonalise this new Hamiltonian $H'(\delta)$ by using the same set of basis wave functions as used for $H$.  As we strengthen the coupling constant $\delta$ from the physical value, $\delta=0$, the eigen-energy of any resonance state gets lower and finally the resonance state changes into a bound state. On the other hand continuum states show almost no change in their eigen-energies for the change of $\delta$.

\begin{figure}[htbp]
\begin{center}
\includegraphics[scale=1.]{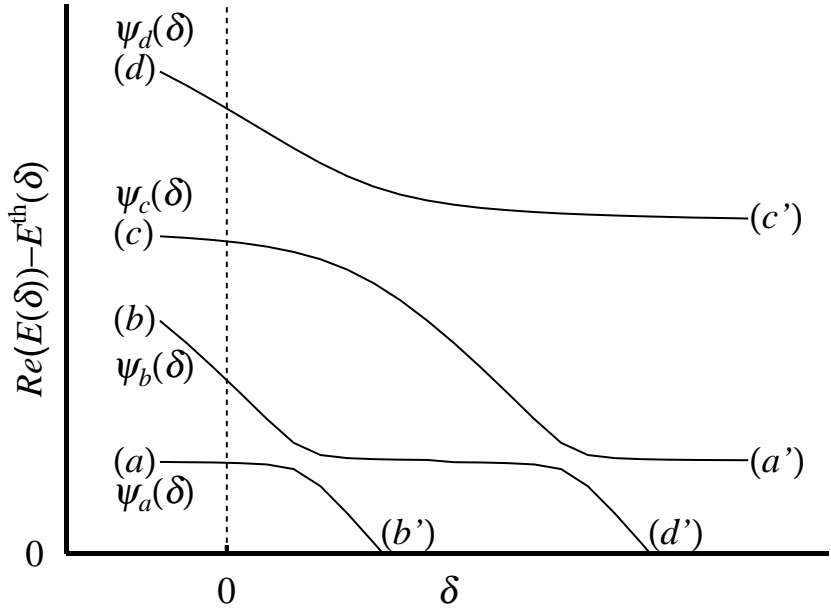}
\caption{Eigen-energies of Hamiltonian (\ref{hamil}) are schematically shown as functions of $\delta$. There are three crossing points near which the mixing of continuum and resonance components occurs. See the text for the detailed explanation.}
\label{fig:1}
\end{center}
\end{figure}

 These behaviors of the eigen-energies as functions of $\delta$ are shown schematically in Fig. \ref{fig:1}.  In this figure, the curve with label (b) around $\delta=0$ shows the eigen-energy of a resonance state denoted as $\psi_{\rm b}(\delta)$ and goes down as $\delta$ increases.  This curve (b) then crosses with the curve (a) which is flat around $\delta=0$ and is the eigen-energy of a continuum state denoted as $\psi_{\rm a}(\delta)$. After the closest approach of the two curves (b) and (a), the curve with label (b') which is the smooth continuation of the curve (a) and hence is disconnected with the curve (b) represents the eigen-energy of the resonance state $\psi_{\rm b}(\delta)$.  Thus, as usual, we say that the curves (b) and (a) have a crossing. The eigen-energy of the continuum state $\psi_{\rm a}(\delta)$ which is first represented by the curve (a) continues to be almost flat except near crossing points with resonance curves and converts into the curve (a'). In the case of the resonance state $\psi_{\rm b}(\delta)$ at $\delta=0$, we can regard that this state contains a negligible amount of mixture of continuum state component because this state is located far from the crossing points with continuum states.

 However, in the case of the resonance state $\psi_{\rm d}(\delta)$ on the resonance state curve (d) in Fig. \ref{fig:1}, its crossing point with the continuum state curve (c) is very close to $\psi_{\rm d}(\delta)$ at $\delta=0$. Therefore $\psi_{\rm d}(\delta)$ at $\delta=0$ should contain sizable mixture of the continuum state.  This kind of situation is usual for a resonance state with broad width, and the resonance wave function obtained by the bound state approximation is not so much trustworthy because of sizable mixture of the continuum state. 

In order to resolve this difficulty, we note the utility of the continuum state wave functions obtained by the bound state approximation in the neighborhood of the resonance state, because they should contain sizable mixture of the resonance state component. In our present simple example of Fig. \ref{fig:1}, we notice the utility of the continuum state wave function $\psi_{\rm c}(\delta)$.  We can express both states,  $\psi_{\rm d}(\delta)$ and $\psi_{\rm c}(\delta)$, as linear combinations of pure resonance state $\Psi_{\rm d}(\delta)$ and pure continuum state $\Psi_{\rm c}(\delta)$; 
\begin{equation}
 \psi_{\rm d}(\delta) = \alpha \Psi_{\rm d}(\delta) + 
  \beta \Psi_{\rm c}(\delta), \quad  
 \psi_{\rm c}(\delta) = -\beta \Psi_{\rm d}(\delta) + 
  \alpha \Psi_{\rm c}(\delta). 
\end{equation}
 The extraction method of the pure resonance component $\Psi_{\rm d}(\delta)$ which we propose is obtained by noticing that the density extension of a resonance state is far smaller than those of continuum states.  Namely we can determine $\alpha$ and $\beta$ from the requirement, 
\begin{eqnarray}
 && \langle \Psi_{\rm d}(\delta) | \sum_{j=1}^A r_j^2 | \Psi_{\rm d}(\delta) 
    \rangle = {\rm smallest}, \\
 && \Psi_{\rm d}(\delta) = \alpha \psi_{\rm d}(\delta) - \beta 
    \psi_{\rm c}(\delta),  
\end{eqnarray} 
 where $A$ is the mass number.  This requirement is satisfied by diagonalizing the operator $\sum_{j=1}^A r_j^2$ by the basis states, $\psi_{\rm d}(\delta)$ and $\psi_{\rm c}(\delta)$, and by choosing $(\alpha, -\beta)$ to be the eigen-vector belonging to the smallest eigen-value.

 When the resonance component is distributed in several energy eigen-states, we need to diagonalize the operator $\sum_{j=1}^A r_j^2$ by such several energy eigen-states. When the energy eigen-states contain two resonance components, the diagonalization of the operator $\sum_{j=1}^A r_j^2$ gives us two small eigen-values.  Two resonance states are obtained by diagonalizing the Hamiltonian $H'(\delta)$ with two wave functions which give the two smallest eigen-values of the operator $\sum_{j=1}^A r_j^2$.

We are now able to calculate approximate values of resonance energies along the resonance energy curve. If a point on the resonance state curve is out of the influence of any crossing points with continuum state curves, we can regard that the energy of the point is just the resonance energy.  On the other hand, if a point on the resonance state curve is located within the influence of some near-by crossing point, we extract resonance wave function 
 by the procedure explained in the previous subsection and then calculate the expectation value of the Hamiltonian $H'(\delta)$ with the extracted wave function which is just the resonance energy at the point.

\subsection{Beyond the ACCC method}\label{newaccc}

 In the ACCC method, the complex eigen-energy $E(\delta)$ of the Hamiltonian $H'(\delta)$ is approximated by Pad\'e rational function, 
\begin{eqnarray}
 && k_{[N,M]} = i \frac{c_0+c_1 x + c_2 x^2 + \cdots + c_N x^N}
    {1 + d_1 x + d_2 x^2 + \cdots + d_M x^M} \label{pade}
    = i \frac{\sum_{j=0}^N c_j x^j}{1+ \sum_{j=1}^M d_j x^j}, \\
 && E(\delta) - E_{\rm th}(\delta) =  k_{[N,M]}^2, \label{comp}
\end{eqnarray} 
 where $E_{\rm th}(\delta)$ is the threshold energy, $x = \sqrt{\delta - \delta_0}$, and $\delta_0$ is determined by the reqirement $E(\delta_0) = E_{\rm th}(\delta_0)$. $c_0$ is reserved to take care of the possible errors in determining $\delta_0$ in practical calculations, though it has to be zero in principle. The $(N + M)$ coefficients, $c_1 \sim c_N, \quad d_1 \sim d_M$, are determined by calculating $E(\delta) - E_{\rm th}(\delta)$ for $(N + M)$ different values of $\delta$ in the bound state region, $\delta > \delta_0$, namely the region where $E(\delta) - E_{\rm th}(\delta) < 0$.

 What we want to know is the complex eigen-energy $E(\delta)$ at $\delta=0$. However, in the ACCC method we have to use the information in the bound state region where $\delta$ is larger than $\delta_0$ and is far from $\delta=0$.  It is the reason why we need high accuracy in the calculation with ACCC.

 As we explained in the previous subsection we are now able to calculate approximate values of resonance energies along the resonance state curve. The knowledge of the resonance energies enables us to determine the $(N + M)$ coefficients, $c_1 \sim c_N, d_1 \sim d_M$, as we explain below. Once the Pad\'e rational function is obtained, we can calculate the resonance widths by using it. The merit of our method is that we can construct the Pad\'e rational function not only by using the bound state information with $\delta > \delta_0$ but by using the information in the resonance state region where $\delta$ is smaller than $\delta_0$ and is near $\delta=0$.

 The explicit procedure to determine the $(N + M)$ coefficients, $c_1 \sim c_N, d_1 \sim d_M$, is as follows.  We calculate the resonance energies ${\cal R}e E(\delta)$ for $L$ different values of $\delta$, $\delta_m, m = 1 \sim L$. The constant $L$ is an arbitrary integer. Then we have the following relation,
\begin{eqnarray}
 & &2\Big({\cal R}e E(\delta_m) - E_{\rm th}(\delta_m)\Big) = (k_{[N,M]})^2 + 
  (k_{[N,M]}^*)^2  \nonumber \\
  & & \quad \quad \quad = - \Big(\frac{\sum_{j=0}^N c_j x_m^j}
     {1+ \sum_{j=1}^M d_j x_m^j}\Big)^2 - 
     \Big(\frac{\sum_{j=0}^N c_j (x_m^*)^j}{1+ \sum_{j=1}^M d_j (x_m^*)^j}\Big)^2, 
     \label{eq:simul} \\
  & &  \quad  x_m =  \sqrt{\delta_m - \delta_0}\ \ (\delta_m \geq \delta_0), \nonumber \\ 
  & &  \quad  x_m = -i \sqrt{\delta_0 - \delta_m}\ \ (\delta_m < \delta_0).
\end{eqnarray}  
 The determination of the $(N + M)$ coefficients, $c_1 \sim c_N, d_1 \sim d_M$, by solving these simultaneous equations can be made by using the following frictional cooling method\ \cite{HMOY,wilet} used widely in the works with AMD (antisymmetrized molecular dynamics) theory\ \cite{ono,kanada},
\begin{eqnarray}
 &&\frac{d}{dt} c_m = - |\mu| \frac{\partial F}{\partial c_m}, \quad 
   \frac{d}{dt} d_m = - |\mu| \frac{\partial F}{\partial d_m}, 
   \label{eq:cool} \\
 &&F = \sum_{m=1}^{L} \Big[ 2\Big({\cal R}e E(\delta_m) - 
   E_{\rm th}(\delta_m)\Big) + \Big(\frac{\sum_{j=0}^N c_j x_m^j}
   {1+ \sum_{j=1}^M d_j x_m^j}\Big)^2 \nonumber \\
 &&\quad \quad \quad  + \Big(\frac{\sum_{j=0}^N c_j (x_m^*)^j}
   {1+ \sum_{j=1}^M d_j (x_m^*)^j}\Big)^2 \Big]^2. \label{f}
\end{eqnarray}  
 Here the $(N + M)$ coefficients, $c_1 \sim c_N, d_1 \sim d_M$, are 
 regarded as being functions of time $t$ and $\mu$ is an arbitrary non-zero real number. It is easy to show that $F$ decreases as time develops, $(d/dt)F < 0$.  Hence after sufficient time steps, we obtain the minimum value of $F$, which is zero.  Needless to say, $F=0$ is equivalent to that the simultaneous equations of Eq.(\ref{eq:simul}) are satisfied. It is to be noted that the initial values of $c_1 \sim c_N, d_1 \sim d_M$ for solving the frictional cooling equation of Eq.(\ref{eq:cool}) are arbitrary.

\section{Application to the $2_2^+$ state of $^{12}$C}\label{appli}

In the previous paper by present authors \cite{funakib}, the $2_2^+$ state of $^{12}$C was studied with the following type of wave function,
\begin{eqnarray}
&&{\widehat \Phi}^{J=2}_{3\alpha}(\vc{\beta}) ={\widehat P}_{J=2} {\cal A} \Big[\prod_{i=1}^3 \exp \Big\{- \Big( \frac{2}{B_x^2} (X_{ix} - X_{Gx})^2+\frac{2}{B_y^2} (X_{iy} - X_{Gy})^2 \nonumber \\ 
&& \hspace{2cm} +\frac{2}{B_z^2} (X_{iz} - X_{Gz})^2  \Big) \Big\} \phi(\alpha_i) \Big]. \label{eq:1}
\end{eqnarray}
Here, $B_k^2 = b^2 + 2\beta_k^2$, ($k=x,y,z$). $\phi ({\alpha}_i) \propto \exp [-(1/2b^2) \sum_j^4(\vc{r}_{j(i)}-{\bf X}_i)^2]$ with $\vc{X}_i =(1/4) \sum_j^4 \vc{r}_{j(i)}$ and $j(i)=4(i-1)+j$ is the internal wave function of the $i$-th alpha cluster, $\vc{X}_G$ is the total center-of-mass coordinate and ${\widehat P}_{J=2}$ is the projection operator of the angular momentum onto $J=2$.  The wave function of Eq.(\ref{eq:1}) is just the $3\alpha$ case of our general $n\alpha$ Bose-condensed wave function proposed recently \cite{THSR,funakia,funaki}. The eigen-energies in bound state approximation are obtained by solving the following Hill-Wheeler equation, 
\begin{equation}
\sum_{\vc{\beta^\prime}}
 \big\langle {\widehat \Phi}^{J=2}_{3\alpha}(\vc{\beta}) \big| (H^\prime(\delta)-E_{\lambda}(\delta)) \big| {\widehat \Phi}^{J=2}_{3\alpha}(\vc{\beta^\prime}) \big\rangle f^{J=2}_\lambda (\vc{\beta^\prime}) =0 .  \label{eq:2}
\end{equation}
The solution of the above equation is given as,
\begin{equation}
\Psi^{J=2}_\lambda (\delta) = \sum_{\vc{\beta}} f^{J=2}_\lambda (\vc{\beta},\delta) {\widehat \Phi}^{J=2}_{3\alpha}(\vc{\beta}). \label{eq:3}
\end{equation}
The microscopic Hamiltonian is expressed as
\begin{equation}
H=T-T_G+V_N+V_C.
\end{equation}
Here $T$ stands for the kinetic energy, $T_G$ the total center-of-mass kinetic energy, $V_C$ the Coulomb energy between protons, and $V_N$ the effective two-nucleon force. In this paper, the Volkov No.2 force \cite{volkov} with Majorana exchange mixture $M=0.59$ is adopted as $V_N$ and the following form as the pseudo potential $V$ in $H^\prime(\delta)=H+\delta \times V$, 
\begin{equation}
V=-80[{\rm MeV}] \exp\Big(-\frac{r_{ij}^2}{(2.5{\rm fm})^2}\Big).
\end{equation}
We should note that the same Hamiltonian $H^\prime(\delta)$ as in Ref.\cite{funakib} is adopted in this paper, so that we can check whether our method works well or not by comparing the results of the new ACCC with those of the original ACCC used in Ref.\cite{funakib}.

\begin{figure}[htbp]
\begin{center}
\includegraphics[scale=1.0]{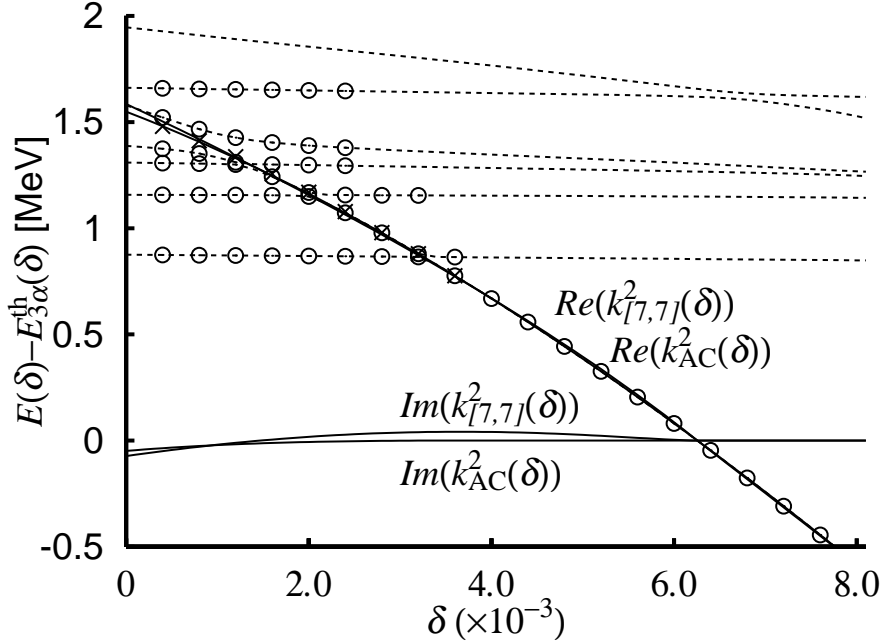}
\caption{The comparison between the complex energy functions which are obtained by the new ACCC method and the original one. These functions are denoted by solid curves. The difference between the real parts of the complex energy functions obtained by both methods is almost invisible. Dotted curves show the energy-eigen values which are the solutions of Hill-Wheeler equation Eq. (\ref{eq:2}). Cross points denote the energy levels obtained by the extraction method explained in \S \ref{extraction}, where the energy-eigen values denoted by open circles in $\delta < 0.004$ are used. The energy levels denoted by open circles in $\delta > 0.004$ and cross points are adopted to construct the complex energy function expressed by the Pad\'e approximant.}
\label{fig:2}
\end{center}
\end{figure}

In Fig. \ref{fig:2}, the complex energy function $k_{[7,7]}^2(\delta)$ for $N=M=7$ calculated by using the extended version of ACCC method is drawn, together with that of the original ACCC method which is denoted as $k_{\rm AC}^2(\delta)$. The $k_{\rm AC}^2(\delta)$ was carefully determined by using the Pad\'e approximant of $N=M=8$ in Ref. \cite{funakib}. Dotted curves denote the eigen-energy levels obtained by solving the Hill-Wheeler equation Eq. (\ref{eq:2}). The eigen-energy levels which are used in order to determine the coefficients, $c_1 \sim c_7$, $d_1 \sim d_7$ of the Pad\'e rational function Eq. (\ref{pade}) are denoted by open circles. We superpose the denoted levels by open circles in order to derive the correct resonance states by the extended version of bound state approximation explained in \S \ref{extraction}, which we simply call the extraction method in the following. This is done at each position of $\delta$ which is less than $0.004$ in the resonance region. The obtained energy levels by the extraction method are denoted by crosses.

We see that the energy levels which are given by the extraction method and denoted by crosses are almost exactly on the trajectory ${\cal R}e( k_{\rm AC}^2(\delta))$. This means that the energy levels given by the extraction method well reproduce the resonance energies predicted by the original ACCC method in the wide region of $\delta$. It is to be noted that since in this $\delta$ region less than $0.004$ the influence of the near-by crossing point is not negligible, without the extraction method the eigen-energy of usual bound state approximation cannot be on the trajectory ${\cal R}e( k_{\rm AC}^2(\delta))$. Thus we can say that our extraction method works effectively and therefore, the obtained wave function is more or less reasonable. We have made a detailed analysis of thus obtained wave function of the $2_2^+$ state in Ref.\cite{funakib}.

Reflecting the fact that all of the energy levels in the new ACCC method, i.e. the levels denoted by crosses for $\delta < 0.004$ and by open circles for $\delta \geq 0.004$, are on the trajectory ${\cal R}e( k_{\rm AC}^2(\delta))$ the real part of the complex energy function ${\cal R}e( k_{[7,7]}^2(\delta))$ coincides with that of the original ACCC method ${\cal R}e( k_{\rm AC}^2(\delta))$. The coincidence of the imaginary part ${\cal I}m( k_{[7,7]}^2(\delta))$ with ${\cal I}m( k_{\rm AC}^2(\delta))$ is not so good as the real part. Nevertheless, we can say the agreement of both trajectories is nice by taking into account the fact that the new and original ACCC method are usually followed by numerical errors to some extent. The calculated resonance energy ${\cal R}e( k_{[7,7]}^2(\delta=0))=1.59$ MeV, which is mesured from the $3\alpha$ threshold, and the width $-2{\cal I}m( k_{[7,7]}^2(\delta=0))=0.15$ MeV, while ${\cal R}e( k_{\rm AC}^2(\delta=0))=1.55$ MeV and $-2{\cal I}m( k_{\rm AC}^2(\delta=0))=0.10$ MeV in the original ACCC method. We have checked that the other choices of $[N, M]$ of the Pad\'e rational function, namely $N=M=6, 8$ resulted in almost the same trajectories for both real and imaginary parts. Incidentally, the observed binding energy and width for the $2_2^+$ state in $^{12}$C are $2.6\pm 0.3$ MeV measured from the $3\alpha$ threshold and $1.0\pm 0.3$ MeV, respectively. It may be said that the theoretical values of the width is much smaller than the experimental value. We discussed this point in the previous work \cite{funakib}, where it was shown that this occurs because the theoretical binding energy is lower than the experimental one and this makes the decay penetrability smaller. See Ref. \cite{funakib} for further detailed discussion.

We examined other input eigen-energy values or other number of the adopted eigen-energy levels in determining the coefficients of the Pad\'e approximant. Except for the case that the eigen-energy values in bound state region are not used, the resultant complex energy function is not ill-behaved but qualitatively unique. The merit of the new ACCC method is that arbitrarily large number of eigen-energies can be used to determine the $(N+M)$ coefficients of Pad\'e rational function so far as the valid numerical accuracy is assured in the frictional cooling method. This should be compared with the original ACCC method which allows just ($N+M$) eigen-energies to be used. Owing to this advantage, we can obtain the correct trajectory of the real part of complex energy function with almost no ambiguity in the wide range of $\delta$ up to $\delta=0$, and therefore the imaginary part can necessarily result in almost unique trajectory at least in the adopted $\delta$ region. It is well known that in the original ACCC method the divergence or ill-behavior easily occur in the obtained energy function unless sufficiently high numerical accuracy is assured.

As mentioned above, in the new ACCC method we can construct the Pad\'e rational function not only by using the bound state information with $\delta > \delta_0$ but also by using the information in the resonance state region where $\delta$ is smaller than $\delta_0$ and is near $\delta=0$. This allows us to determine the coefficients of Pad\'e rational function with less numerical accuracy. In fact, we stop the time development of $F$ of Eq. (\ref{f}) before it reaches around $10^{-3}$ in practical calculations. On the contrary, in the original ACCC method only the eigen-energy values in the bound state region of $\delta \geq \delta_0$ are used, and hence high numerical accuracy is needed for the extrapolation of the Pad\'e rational function up to $\delta=0$ which is far from the bound state region. 

\section{Conclusion}\label{conc}

We proposed a new method beyond the ACCC method in which in order to determine the coefficients of Pad\'e approximant to the complex eigen energy we adopt the negative and positive energies which are obtained in bound state approximation or an extended version of bound state approximation to the description of resonances. The determination of the coefficients of Pad\'e approximant was made by the use of the frictional cooling method which has been widely used in the works with AMD theory. The extended version of bound state approximation was already utilized in the previous work by present authors, and here we gave a full explanation of the new method. The new method enables us to extract the resonance component from the energy eigen-states in the bound state approximation which consist of the mixture of the resonance state and the continuum state components. 

We investigated the validity of the new method by the application to the second $2^+$ state of $^{12}$C which was discussed with the original ACCC method in our previous paper. We showed that the resonance states which were obtained in the extended version of bound state approximation correctly give the resonance energies which are determined by the use of the original ACCC method. The use of energy levels in the wide region of coupling constant which includes both resonance state and bound state was found to lead to the unique construction of the Pad\'e rational function, which qualitatively well agrees with that of the original ACCC method. Thus we obtained similar values of energy and width to those given in the original ACCC method. The number of adopted eigen-energy states to construct the Pad\'e rational function can be taken irrespectively of the dimension of the function, i.e. $(N+M)$, while in the original ACCC method the number is fixed to $(N+M)$. It should be noted that this new method needs much less numerical accuracy compared with the original one in which eigen value problem must be solved with extremely high numerical precision. We can conclude that our new method has several advantages over the original one and can be applied directly to resonance problems whenever the original ACCC method is applicable.


\section*{Acknowledgements}
 Authors highly appreciate many helpful comments and words of advice of Prof. P. Schuck and Prof. G. R\"opke. Valuable comments of Prof. S. Aoyama are also acknowledged. This work was partially performed in the Research Project for Study of Unstable Nuclei from Nuclear Cluster Aspects sponsored by Institute of Physical and Chemical Research (RIKEN), and is supported by the Ministry of Education, Science, Sports and Culture, Grant-in-Aid for JSPS Fellows, (No. 15$\cdot$5511) and by the Grant-in-Aid for the 21st Century COE "Center for Diversity and Universality in Physics" from the Ministry of Education, Culture, Sports, Science and Technology (MEXT) of Japan.

%

\end{document}